\documentclass[runningheads]{llncs}
\usepackage[T1]{fontenc}
\usepackage{graphicx}
\usepackage{amsmath,amssymb,amsfonts}
\usepackage{textcomp}
\usepackage{xcolor}
\usepackage{tabularx}   
\usepackage{cite}
\usepackage{float}
\usepackage{booktabs}
\usepackage{multirow}
\usepackage[hidelinks]{hyperref}
\usepackage{threeparttable}

\begin{document}

\title{Agentic AI in Industry: Adoption Level and Deployment Barriers}

\author{Spyridon Alvanakis Apostolou\inst{1}\thanks{Corresponding author.} \and
Jan Bosch\inst{1,2} \and
Helena Holmström Olsson\inst{3}}

\institute{Chalmers University of Technology, Department of Computer Science and Engineering, Gothenburg, Sweden
\email{spyalv@chalmers.se} \and
Eindhoven University of Technology, Department of Mathematics and Computer Science, Eindhoven, Netherlands \email{jan.bosch@chalmers.se} \and
Malmö University, Department of Computer Science and Media Technology, Malmö, Sweden \email{helena.holmstrom.olsson@mau.se}}

\maketitle             

\begin{abstract}

Agentic AI is entering software engineering workflows, but empirical evidence on its transition from experimental capability to production use remains limited. We report a qualitative interview study with 16 practitioners from 12 companies, using a six-level maturity framework as an analytical lens. Reported production practices corresponded to Levels 1--3, while participants in four companies reported experimental capabilities beyond production-integrated use. Across the cases, four previously identified barriers recurred: context management, performance on proprietary content, non-determinism and qualification, and data confidentiality. We synthesize their interaction as a capability–deployment verification gap structured by two interdependent dimensions: information asymmetry and qualification absence. The study thereby characterizes reported adoption practices and explains what constrains further agentic automation in the represented industrial contexts.

\end{abstract}

\section{Introduction}
The emergence of LLM-based software engineering tools has accelerated automation within software product development. Many industrial implementations remain assistant-oriented, however, newer agentic integrations extend this paradigm toward task completion, tool use, iterative refinement, and autonomous decision-making. These capabilities move automation within the Software Development Life Cycle (SDLC) beyond code-generation assistance toward goal-directed systems \cite{he_llm-based_2024}. This transition raises two questions: what forms of agentic integration are currently used in industrial software engineering, and what prevents further advancement of this type of automation?

Despite the broad interest in LLM-based implementations, studies focusing on industrial adoption remain sparse. We therefore conducted an interview study with 16 practitioners from 12 companies of varying size and domain. An adaptive semi-structured questionnaire, guided by a six-level maturity framework, elicited the production practices and experimental capabilities visible to participants.

Using the maturity framework as an analytical lens, we make three contributions. First, we provide a bounded empirical characterization of reported production practices and experimental capabilities corresponding to Levels 1--3. Second, we show how four previously identified barriers (context management, performance on proprietary content, non-determinism and qualification, and data confidentiality) interact in the studied contexts, and document mitigation strategies in four companies where experimental capabilities exceeded production-integrated practice. Third, we synthesize these observations into a capability--deployment verification gap structured by two interdependent dimensions: information asymmetry and qualification absence.

\section{Background and Related Work} \label{background}

\subsection{AI in SDLC and Shift to Automation} \label{AIforAutomation}

We define LLM-based software engineering tools as tools whose core capabilities rely on large language models. Agentic capabilities add goal-directed behavior through planning, tool use, context retrieval, execution, iterative refinement, automated decision-making, or coordination~\cite{he_llm-based_2024}. Agentic-capable tools are therefore a subset of LLM-based tools, while agentic integration denotes their use to perform SDLC tasks automatically. The maturity trajectory covers LLM-based integration broadly. During the interviews, we assessed whether implementations exhibited the capabilities above. The framework is thus an integration-maturity lens for progression toward agentic automation, not a taxonomy in which every level is agentic.

Assistant-oriented tools already support code completion, conversation, and review, while agentic systems extend this support through planning and multi-step execution. Studies report measurable productivity improvements in tasks supported by both assistant-oriented~\cite{liang_large-scale_2023, stray_developer_2026} and agentic~\cite{sawant_agentic_2025} systems, motivating further automation. Greater task ownership also amplifies friction already visible in LLM-based tools: systems may omit aspects of developer intent, vary across equivalent prompts, hallucinate, or lose relevant information as context grows~\cite{yujia_security_2025, mastropaolo_robustness_2023}. These challenges motivate our focus on reliable information access and output verification when integrating demonstrated capabilities into production workflows.

\subsection{Related Work}

Empirical evidence on organizational adoption remains limited. Across industries, Hughes et al.~\cite{hughes_ai_2025} identify legacy compatibility, accountability, and missing governance for autonomous operation. In software engineering, Akbar et al. find adoption concentrated in code generation and bug detection, with tool maturity and human skills limiting broader use~\cite{akbar_agentic_2025}. Work in embedded software highlights determinism, traceability, and regulatory compliance in safety-critical settings~\cite{sun_agentic_2026}, while practitioner interviews by Ferino et al. connect productivity gains with concerns about code quality, privacy, and the continuing need for human oversight~\cite{ferino_walking_2025}.

Collectively, these studies identify barriers involving legacy and proprietary systems, context limitations, non-determinism, confidentiality, governance, and human oversight. They provide less evidence, however, about how these barriers interact when experimental agentic capabilities cannot be integrated into production, or how organizations mitigate that mismatch. This boundary motivates our analysis.

\section{Methodology} \label{methods}

\subsection{Research Design and Maturity Framework}

We conducted an exploratory qualitative interview study and used cross-case synthesis~\cite{yin_case_2009} to examine production-integrated practices, experimental capabilities, challenges, and mitigation strategies within the organizational contexts described by participants.

The interview design used Bosch and Olsson's framework for AI-driven organizational evolution, which was inductively derived from practitioner interviews~\cite{bosch_towards_2026}. We adopted its five levels and added Level 0 to capture individual use preceding organizational provision, Table~\ref{tab1:matassess} summarizes the resulting framework. The terminology for agentic capabilities is in Section~\ref{AIforAutomation}.

The framework served two purposes: it routed interviews to level-specific questions and provided a common vocabulary for comparing practices. For each company, the assigned level was the highest production-integrated practice reported; practices at other levels could coexist across tasks, while non-integrated capabilities were recorded separately as experimental.

\subsection{Participants, Data Collection, and Analysis}\label{PartSelect}

We used purposive sampling \cite{patton_qualitative_2014} to recruit practitioners whose roles provided relevant visibility into integrated LLM-based tools and related adoption decisions. Their accounts should not be interpreted as exhaustive descriptions of organization-wide practice, particularly in large organizations where tools and practices may vary across units. The 16 participants represented 12 companies spanning domains, regulatory environments, and sizes: three small (fewer than 100 employees), two medium (100--1000), and seven large (more than 1000). We considered a company regulated when external rules constrained its development process or tooling. All participants consented to anonymous participation, and the transcripts are retained confidentially by the research team. Detailed sample characteristics appear in  \href{https://doi.org/10.5281/zenodo.22962280}{\underline{Supplementary Table S1}}.

The semi-structured interviews lasted more than 45 minutes on average and covered four areas: 1) participant and organizational context, 2) maturity assessment, 3) a level-specific examination of practices, challenges, mitigations, and unmet needs, and 4) future opportunities for agentic AI. The maturity assessment routed each interview to relevant follow-up questions rather than imposing an identical question set across contexts. The complete \href{https://doi.org/10.5281/zenodo.22962280}{\underline{interview protocol}} is available online .

\begin{table}[t]
    \caption{Agentic AI Maturity Levels}
    \label{tab1:matassess}
    \footnotesize
    \centering
    \begin{tabularx}{\columnwidth}{l l >{\raggedright\arraybackslash}X}
    \toprule
    \textbf{Lvl} & \textbf{Name} & \textbf{Description} \\
    \midrule
    0 & Individual Use     & Personal use of LLM-based tools without org. provision \\
    1 & AI Assistants      & Organizationally provisioned assistance LLM-based tools, including lightweight agentic use \\
    2 & Task Agents        & Deployed agents with bounded ownership of specific SDLC tasks \\
    3 & Collaborative AI   & Human-orchestrated agents that delegate subtasks across SDLC phases \\
    4 & System Builders    & Agents generate complete software from high-level intent \\
    5 & Self-Optimizing AI & Agents autonomously monitor, repair, and regenerate software \\
    \bottomrule
    \end{tabularx}
\end{table}

Each transcript was reviewed and summarized using a common structure: participant context, reported integration level and tools, workflow and experimental practices, and limitations and needs. These summaries were systematically compared across contexts, reported limitations were iteratively grouped and reassessed until the higher-order barrier categories stabilized. Two locally hosted LLMs were used only to flag potential omissions from the summaries: all suggestions were manually checked against the transcripts, 11 contextual additions were accepted, and the researchers retained all analytical decisions. A supporting \href{https://doi.org/10.5281/zenodo.22962280}{\underline{evidence table}} provides traceable interview articulations.

\section{Findings} \label{findings}

\subsection{Reported Production Practices and Limitations}

Participants described highest production-integrated practices corresponding to Level 1 in seven companies, Level 2 in four, and Level 3 in one, none reported production use corresponding to Levels 0, 4, or 5. The distribution by regulatory status is provided in  \href{https://doi.org/10.5281/zenodo.22962280}{\underline{Supplementary Table S2}}.

\textbf{At level 1:} Across seven companies, participants reported assistants for code generation, debugging, documentation, and explanation. Narrow IDE- or CLI-based agentic processes remained ad hoc, requiring context and intent to be re-established for each use. Two participants described complex or niche tasks for which it was \textit{"faster to write code from scratch"} than to evaluate generated code line by line and debug it. Participants also noted fear of losing awareness of the codebase as generation outpaced human review. Other limitations included incorrect or non-deterministic output, fragmented information, weak performance on proprietary content, and on-premises restrictions.

\textbf{At level 2:} Across four companies, practices assigned agents bounded tasks such as documentation generation, pull-request analysis, code review, and log or fault analysis. These applications shared an important property: their outputs were auditable and remained subject to human review, tests, or other guardrails before propagating further. Participants did not report agents autonomously completing a pipeline stage. One participant described agents as tending to \textit{"overengineer"} solutions by generating unnecessary code. Human verification was therefore used to check both technical output and semantic intent. The constraint was sharper in safety-regulated companies: one required two evaluators for code review to reduce errors.

\textbf{At level 3:} The single Level 3 company used multi-agent pipelines to delegate tasks and re-rank retrieved context in client-facing cloud deployments. Errors could still propagate through longer interactions, and outputs required human semantic evaluation. Across the reported levels, the salient distinction was therefore not the presence of capable agents, but whether their information access and outputs could be bounded well enough for production integration.

\subsection{Capability without Deployability and Cross-cutting Barriers} \label{Gap}

Four recurring barriers constrained such integration. \textbf{Context management} (CM) concerned both context-window capacity and the fragmentation of large codebases, documentation, requirements, architecture, and regulations across independently maintained sources. \textbf{Proprietary-content performance} (PC) concerned not only code, but also internal definitions, packages, domain-specific languages, hardware toolchains, and company protocols poorly represented in model training data. \textbf{Non-determinism and qualification incompatibility} (ND) made behavior difficult to bound, while tests and static analysis could not establish semantic correctness or conformance to intent. \textbf{Data confidentiality} (DC) restricted the information that could leave organizational premises and often required smaller local models with lower reported capability than larger cloud-hosted models. 
Table~\ref{tab:barriers} summarizes the barriers associated with the reported practices, grouped rows indicate barriers reported in one or more of the listed companies and do not imply that every barrier was reported in every company.

\begin{table*}[t]
    \caption{Reported production practices, experimental capabilities,
    barriers, and mitigation strategies across the represented organizations.}
    \label{tab:barriers}
    \footnotesize
    \centering
    \setlength{\tabcolsep}{4pt}
    \renewcommand{\arraystretch}{1.08}
    \begin{tabularx}{\textwidth}{@{}
        >{\raggedright\arraybackslash}p{1.35cm}
        >{\raggedright\arraybackslash}p{2.35cm}
        >{\raggedright\arraybackslash}X
        >{\raggedright\arraybackslash}p{1.55cm}
        >{\raggedright\arraybackslash}X@{}}
    \toprule
    \textbf{Prod. level} & \textbf{Production practice} &
    \textbf{Higher-level experiment} & \textbf{Barriers} &
    \textbf{Mitigation and remaining limit} \\
    \midrule
    Level 1 (5 companies) & Assistants & None reported & CM, ND, DC vary &
    Local tools or manual review; no stable workflow \\
    \addlinespace[2pt]
    Level 1 (2 companies) & Assistants & Information retrieval; code-review agent evaluation &
    CM, PC, ND, DC & Restricted tasks; still under evaluation \\
    \midrule
    Level 2 (2 companies) & Task agents & None reported & CM, PC, ND vary &
    Human and formal review; verification cost remains \\
    \addlinespace[2pt]
    Level 2 (2 companies) & Bounded task agents & Risk assessment and test generation; multi-agent review & CM, PC, ND, DC & Hierarchical/Graph-RAG,
    local models, and guardrails remain partial \\
    \midrule
    Level 3 (1 company) &  Multi-agent workflow & None reported & CM, ND &
    Native platform tools; semantic evaluation required \\
    \bottomrule
    \end{tabularx}
\end{table*}

The clearest separation between experimental capability and production-integrated practice appeared in four of twelve companies. Here, \textbf{deployment} denotes integration into an active development workflow. One Level 1 company used agent-assisted retrieval of fragmented product knowledge but avoided AI-assisted embedded-code generation because review and compiler-aware debugging remained difficult. As one participant summarized, \textit{"It is an unfair comparison between the developer and the agents, since the agents do not have all the information."} Another Level 1 company evaluated task-specific code-review agents without integrating them into its workflow.

One safety-regulated Level 2 company illustrates qualification absence. Its experimental multi-agent risk-assessment pipeline decomposed malfunction scenarios, delegated severity assessments, and produced a requirements table, but a human remained responsible solely for the final evaluation to avoid introducing bias into the assessment chain. Safety-related use required documented behavior boundaries that the non-deterministic system could not provide. As a participant explained, \textit{"In safety-related fields, if you want to use a tool you need to know what are the bugs or the weaknesses in that tool to be able to use it in your daily task"}. 

For information management and organization, the same company reported that conventional RAG worked for direct questions over well-structured documentation but not for complex questions spanning requirements, architecture, regulation, and code. Graph-RAG improved cross-source retrieval, yet its knowledge boundaries still had to be predefined for each context, while a cloud-hosted implementation could access only cloud-stored sources. For generated code entering integration and testing, layered static-analysis and testing tools reduced detectable code-level errors, but did not establish semantic correctness or conformance to cross-cutting regulatory intent.

Another Level 2 company used proprietary languages and hardware. Its bounded multi-agent code-review workflows were integrated into active use and reportedly reduced bug-resolution turnaround from days or weeks to hours. Broader end-to-end orchestration remained outside the production workflow. Its hierarchical retrieval approach batched large inputs into a vector database enriched with internal documentation, but keeping that representation synchronized with an evolving codebase created substantial maintenance work. Sandboxed, organizationally hosted local models addressed confidentiality and versioning concerns, but were less capable than larger cloud models, internal adaptation for proprietary languages and packages remained costly and limited. Reproducibility additionally required earlier model versions to remain identifiable and available so that previous results and code states could be reproduced, making continued model availability a contractual concern when external vendors were used.

Human evaluation remains the trusted gate and a deployment bottleneck. All Level 2 companies required a human at least for verification, while one safety-regulated company required two evaluators for code review. One participant emphasized augmentation rather than replacement: automation should reduce cognitive load without eroding the quality gatekeeping required by regulation. This gate becomes harder to scale as generated output grows. The problem extended beyond review: no individual can maintain coherent awareness of a 100,000 ticket backlog, fragmented information across organizational departments, or the full regulatory traceability of a complex product. 

\section{Discussion} \label{discussion}

The individual barriers are not claimed as new: they align with prior reports of legacy and governance constraints, tool maturity, determinism, traceability, privacy, and human oversight~\cite{hughes_ai_2025, akbar_agentic_2025, sun_agentic_2026, ferino_walking_2025}. Our contribution is their empirical synthesis into two interdependent dimensions, \textit{information asymmetry} and \textit{qualification absence}, that explain the separation between experimental capability and production-integrated practice.

\textbf{Information asymmetry.} This dimension connects three empirically distinct barriers: fragmented organizational knowledge, more information than an agent can use at once, and weak model performance on proprietary content. Functionally, each leaves the agent without some of the context that determines whether its output is appropriate. Data confidentiality is analytically distinct as an infrastructural constraint, yet it can intensify the asymmetry by restricting which sources a model may access. Information asymmetry is therefore not simply a context-window problem, but the absence of an accessible, current, and task-relevant representation of organizational knowledge.

\textbf{Qualification absence.} Providing more information does not establish that an output satisfies the intended task. The verification mechanisms reported by participants detect particular faults, but do not fully establish semantic correctness or conformity to requirements, architectural constraints, and development intent. Human qualification therefore exists and remains the trusted gate. What is absent in the represented contexts is a scalable automated counterpart for tasks whose acceptance criteria cannot be reduced to execution, tests, or static rules. 
% This matters because the volume of generated output and relevant product information can grow faster than individuals can review or track coherently.

Together, the two dimensions constitute the \textit{capability--deployment verification gap}. Automated qualification depends on the completeness and provenance of the information against which an output is judged, while improved information access remains insufficient without a means to establish that the resulting action is acceptable.

The single Level 3 company provides a contrast: a consolidated cloud platform and limited legacy burden supported a production multi-agent workflow, but outputs still required human semantic evaluation. Most reported uses were code-adjacent, where compilation, tests, static analysis, and logs provide partial executable oracles consistent with the concentration observed by Akbar et al.~\cite{akbar_agentic_2025}. Requirements engineering and architecture depend more heavily on distributed, multi-modal knowledge and offer weaker executable checks, intensifying both dimensions. Moving beyond bounded code-related tasks therefore requires mechanisms that connect current product knowledge to intent-aware qualification, so that the evidence needed to trust an agent's output scales with it.

\section{Threats to Validity} \label{threats}

\textbf{Internal validity.} This exploratory study relies on participant reports rather than direct observation. Adaptive, level-specific interview routes improved detail but reduced uniform cross-level comparison. \textbf{Construct validity.} Participants had relevant visibility, but their accounts may miss organization-wide variation, particularly in the four large companies represented by one participant. With no reported production practices at Levels 4--5, we cannot assess the framework or barriers at those levels.  \textbf{External validity.} Purposive recruitment through professional networks favored AI-aware participants, while seven of the twelve companies were large, limiting transferability to less AI-active and smaller organizations.

\section{Conclusion \& Future Directions} \label{conclusion}

Based on 16 interviews across 12 companies, we used a six-level maturity framework to describe reported practices. In four companies, experimental capability exceeded production-integrated practice. We interpret the recurring barriers as a capability--deployment verification gap structured by information asymmetry and qualification absence. Reported mitigations reduced but did not close these constraints, leaving human review as gate that becomes harder to scale as generated output grows. This gap defines a \textbf{future research direction}: unified qualification and information-navigation mechanisms that connect current product knowledge and its provenance to intent, maintainability, architectural dependencies, and regulatory constraints. Such mechanisms cannot eliminate probabilistic error, but may narrow it, and enable broader (more auditable and deployable) agentic integration.

\subsubsection*{Acknowledgements and Disclosure of Interests.} We thank the Software Center companies for partly funding this research work. The authors have no competing interests to declare that are relevant to the content of this article.

% \subsubsection*{Acknowledgements} We thank the Software Center companies for partly funding this research work.

% \subsubsection*{Disclosure of Interests.}
% The authors have no competing interests to declare that are relevant to the content of this article.

\begingroup
\bibliographystyle{splncs04}
\bibliography{InterviewStudy}
\endgroup

\end{document}